\documentclass[showpacs,10pt,twocolumn,prb]{revtex4-1}

\usepackage{amsmath}
\usepackage{amssymb}
\usepackage{graphicx}
\usepackage{amssymb}
\usepackage{graphics}
\usepackage{epsfig}
\usepackage{CJK}
\usepackage{color}

\begin{document}

\begin{CJK*}{GBK}{Song}
\title{Critical behavior and magnetocaloric effect in Mn$_3$Si$_2$Te$_6$}
\author{Yu Liu and C. Petrovic}
\affiliation{Condensed Matter Physics and Materials Science Department, Brookhaven National Laboratory, Upton, New York 11973, USA}
\date{\today}

\begin{abstract}
The critical properties and magnetocaloric effect of semiconducting ferrimagnet Mn$_3$Si$_2$Te$_6$ single crystals have been investigated by bulk magnetization and heat capacity around $T_c$. Critical exponents $\beta = 0.41\pm0.01$ with a critical temperature $T_c = 74.18\pm0.08$ K and $\gamma = 1.21\pm0.02$ with $T_c = 74.35\pm0.05$ K are deduced by the Kouvel-Fisher plot, whereas $\delta = 4.29\pm0.05(3.40\pm0.02)$ is obtained by a critical isotherm analysis at $T = 74(75)$ K. The magnetic exchange distance is found to decay as $J(r)\approx r^{-4.79}$, which lies between the mean-field and 3D Heisenberg models. Moreover, the magnetic entropy change $-\Delta S_M$ features a maximum at $T_c$, i.e., $-\Delta S_M^{max} \sim$ 2.53(1.67) J kg$^{-1}$ K$^{-1}$ with in-plane(out-of-plane) field change of 5 T, confirming large magnetic anisotropy. The heat capacity measurement further gives $-\Delta S_M^{max}$ $\sim$ 2.94 J kg$^{-1}$ K$^{-1}$ and the corresponding adiabatic temperature change $\Delta T_{ad}$ $\sim$ 1.14 K with out-of-plane field change of 9 T.
\end{abstract}
\maketitle
\end{CJK*}

\section{INTRODUCTION}

Layered intrinsically ferromagnetic (FM) semiconductors hold great promise for both fundamental physics and applications in spintronic devices.\cite{McGuire0, McGuire, Huang, Gong, Seyler} CrI$_3$ has recently attracted much attention since the long-range magnetism persists in monolayer with $T_c$ of 45 K.\cite{Huang} Intriguingly, the magnetism in CrI$_3$ is layer-dependent, from FM in monolayer, to antiferromagnetic (AFM) in bilayer, and back to FM in trilayer.\cite{Huang} It can further be controlled by electrostatic doping, providing great opportunities for designing magneto-optoelectronic devices.\cite{Jiang, Huang1}

Ternary Cr$_2$X$_2$Te$_6$ (X = Si, Ge) exhibit FM order below $T_c$ of 32 K for Cr$_2$Si$_2$Te$_6$ and 61 K for Cr$_2$Ge$_2$Te$_6$, respectively,\cite{Ouvrard, Carteaux1, Carteaux2, Casto, Zhang} and also promising candidates for long-range magnetism in nanosheets.\cite{Gong, Lin, Zhuang} Many efforts have been devoted to shed light on the nature of FM in this system.\cite{Li, Chen, Sivadas, Liu, BJLiu, LinGT, YULIU} Multiple domain structure types, self-fitting disks and fine ladder structure within the Y-connected walls, were observed by magnetic force microscopy,\cite{Guo} confirming two-dimensional (2D) long-range magnetism with non-negligible interlayer coupling.\cite{YULIU} Mn$_3$Si$_2$Te$_6$ is a little-studied three-dimensional (3D) analog of Cr$_2$Si$_2$Te$_6$.\cite{Rimet, Vincent, MAY} The Mn$_2$Si$_2$Te$_6$ layer is composed of MnTe$_6$ octahedra that are edge sharing within the $ab$ plane (Mn1 site) and along with Si-Si dimers [Fig. 1(a)], similar to Cr$_2$Si$_2$Te$_6$. However, the layers are connected by filling one-third of Mn atoms at the Mn2 site within interlayer, yielding a composition of Mn$_3$Si$_2$Te$_6$.\cite{MAY} Recent neutron diffraction experiment gives that Mn$_3$Si$_2$Te$_6$ is a ferrimagnet below $T_c \approx 78$ K and the moments lie within the $ab$ plane.\cite{MAY}

In the present work we investigated the critical behavior of Mn$_3$Si$_2$Te$_6$ single crystal by using modified Arrott plot, Kouvel-Fisher plot and critical isotherm analysis, as well as its magnetocaloric effect. Critical exponents $\beta$ = 0.41(1) with $T_c$ = 74.18(8) K, $\gamma$ = 1.21(2) with $T_c$ = 74.35(5) K, and $\delta$ = 4.29(5) at $T$ = 74 K. The magnetic exchange distance is found to decay as $J(r)\approx r^{-4.79}$, which lies between mean-field and 3D Heisenberg models. The rescaled $-\Delta S_M(T,H)$ curves can well collapse onto a universal curve, confirming its nature of second-order.

\section{METHODS}

\subsection{Experimental details}
Single crystals of Mn$_3$Si$_2$Te$_6$ were fabricated by melting stoichiometric mixture of Mn (3N, Alfa Aesar) chip, Si (5N, Alfa Aesar) lump and Te (5N, Alfa Aesar) shot. Starting materials were vacuum-sealed in a quartz tube, heated to 1100 $^\circ$C over 20 h and then cooled to 850 $^\circ$C at a rate of 1 $^\circ$C/h. X-ray diffraction (XRD) data were taken with Cu $K_{\alpha}$ ($\lambda=0.15418$ nm) radiation of a Rigaku Miniflex powder diffractometer. The magnetization and heat capacity were collected in Quantum Design MPMS-XL5 and PPMS-9 systems. The magnetic entropy change $-\Delta S_M$ from the magnetization data was estimated using a Maxwell relation.

\subsection{Scaling analysis}
According to the scaling hypothesis, the second-order phase transition around the Curie point $T_c$ is characterized by a set of interrelated critical exponents $\alpha$, $\beta$, $\gamma$, $\delta$, $\eta$, $\nu$ and a magnetic equation of state.\cite{Stanley} The exponent $\alpha$ can be obtained from specific heat and $\beta$ and $\gamma$ from spontaneous magnetization $M_s$ and inverse initial susceptibility $\chi_0^{-1}$, below and above $T_c$, respectively, while $\delta$ is the critical isotherm exponent. The mathematical definitions of the exponents from magnetization measurement are given below:
\begin{equation}
M_s (T) = M_0(-\varepsilon)^\beta, \varepsilon < 0, T < T_c,
\end{equation}
\begin{equation}
\chi_0^{-1} (T) = (h_0/m_0)\varepsilon^\gamma, \varepsilon > 0, T > T_c,
\end{equation}
\begin{equation}
M = DH^{1/\delta}, T = T_c,
\end{equation}
where $\varepsilon = (T-T_c)/T_c$ is the reduced temperature, and $M_0$, $h_0/m_0$ and $D$ are the critical amplitudes.\cite{Fisher}

The magnetic equation of state in the critical region is expressed as
\begin{equation}
M(H,\varepsilon) = \varepsilon^\beta f_\pm(H/\varepsilon^{\beta+\gamma}),
\end{equation}
where $f_+$ for $T>T_c$ and $f_-$ for $T<T_c$, respectively, are the regular functions. Eq.(4) can be further written in terms of scaled magnetization $m\equiv\varepsilon^{-\beta}M(H,\varepsilon)$ and scaled field $h\equiv\varepsilon^{-(\beta+\gamma)}H$ as
\begin{equation}
m = f_\pm(h).
\end{equation}
This suggests that for true scaling relations and the right choice of $\beta$, $\gamma$, and $\delta$ values, scaled $m$ and $h$ will fall on universal curves above $T_c$ and below $T_c$, respectively.

\section{RESULTS AND DISCUSSIONS}

\begin{figure}
\centerline{\includegraphics[scale=1]{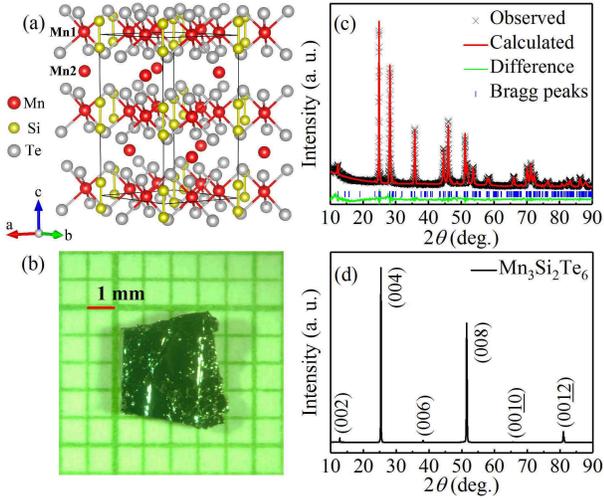}}
\caption{(Color online) (a) Crystal structure and (b) representative single crystal of Mn$_3$Si$_2$Te$_6$. (c) Powder x-ray diffraction (XRD) and (d) single-crystal XRD patterns of Mn$_3$Si$_2$Te$_6$ at room temperature. The vertical tick marks represent Bragg reflections of the $P\bar{3}1c$ space group.}
\label{XRD}
\end{figure}

The powder XRD pattern of Mn$_3$Si$_2$Te$_6$ confirms high purity of the single crystals, in which the observed peaks can be well fitted with the $P\bar{3}1c$ space group [Fig. 1(c)]. The determined lattice parameters $a = 7.046(2)$ {\AA} and $c = 14.278(2)$ {\AA} are very close to the reported values.\cite{Vincent, MAY} In the single-crystal XRD [Fig. 1(d)], only $(00l)$ peaks are detected, indicating that the crystal surface is parallel to the $ab$ plane and perpendicular to the $c$ axis.

Figure 2 presents the temperature dependence of magnetization measured in $H$ = 1 and 50 kOe applied in the $ab$ plane and parallel to the $c$ axis, respectively. The magnetization is nearly isotropic in 50 kOe, however, significant magnetic anisotropy is observed in 1 kOe at low temperatures. The ordered moments lie primarily within the $ab$ plane. An additional upturn well above $T_c$ till to 300 K is clearly seen in zero-field-cooling (ZFC) curve for $\mathbf{H} // \mathbf{ab}$, which may be associated with short-range order or the presence of correlated excitations in the paramagnetic region.\cite{MAY} Isothermal magnetization at $T$ = 5 K [insets in Fig. 2] shows saturation moment of $M_s \approx$ 1.6 $\mu_B$/Mn for $\mathbf{H} // \mathbf{ab}$ and a small FM component for $\mathbf{H} // \mathbf{c}$. No remanent moment for either orientation confirms the crystal of high quality. The $T_c$ can be roughly determined by the minimum of $d\chi/dT$ [Figs. 2(c,d)], i.e., $T_c$ = 75 K for in-plane field and $T_c$ = 77 K for out-of-plane field of 1 kOe, which shifts to $T_c$ = 80 K in an increase field of 50 kOe.\cite{MAY}

\begin{figure}
\centerline{\includegraphics[scale=1]{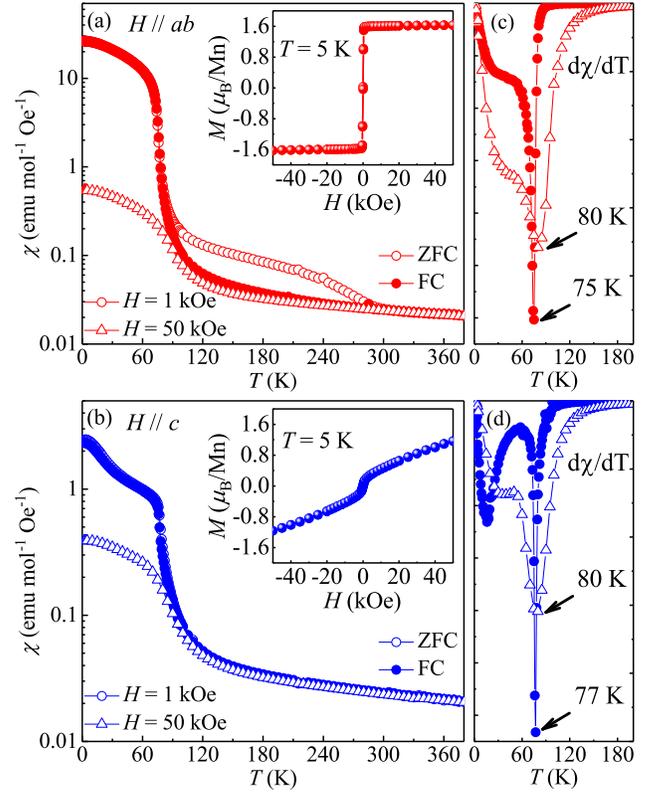}}
\caption{(Color online) Temperature dependence of dc magnetic susceptibility $\chi$ and corresponding $d\chi/dT$ for Mn$_3$Si$_2$Te$_6$ measured in the magnetic field $H$ = 1 and 50 kOe applied (a,c) in the $ab$ plane and (b,d) along the $c$ axis, respectively. Insets: field dependence of magnetization measured at $T$ = 5 K.}
\label{MTH}
\end{figure}

\begin{figure}
\centerline{\includegraphics[scale=1]{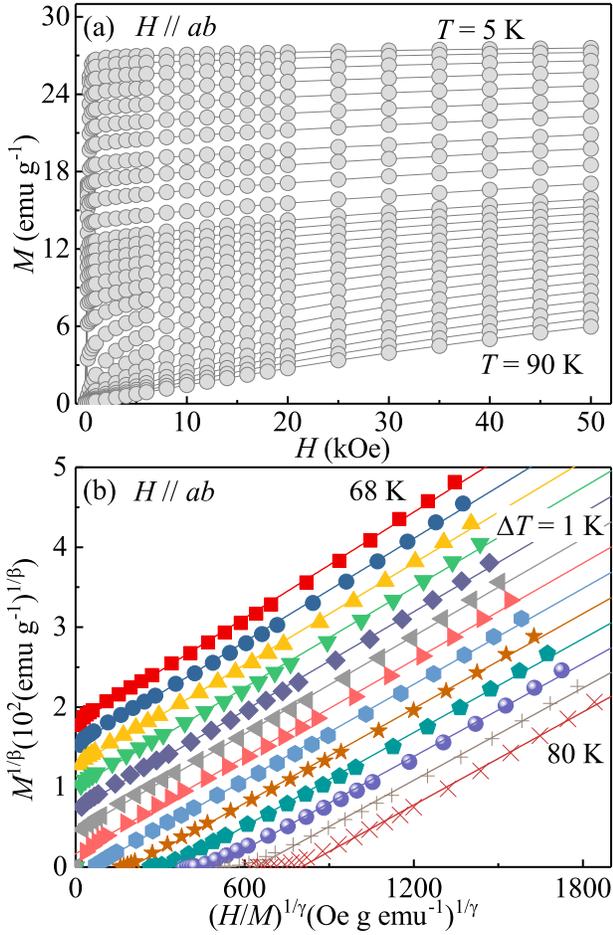}}
\caption{(Color online) (a) Typical initial isothermal magnetization curves measured in $\mathbf{H} // \mathbf{ab}$ from 5 to 90 K for Mn$_3$Si$_2$Te$_6$. (b) the modified Arrott Plot around $T_c$ for the optimum fitting with $\beta = 0.41$ and $\gamma = 1.21$.}
\label{Arrot}
\end{figure}

\begin{figure}
\centerline{\includegraphics[scale=1]{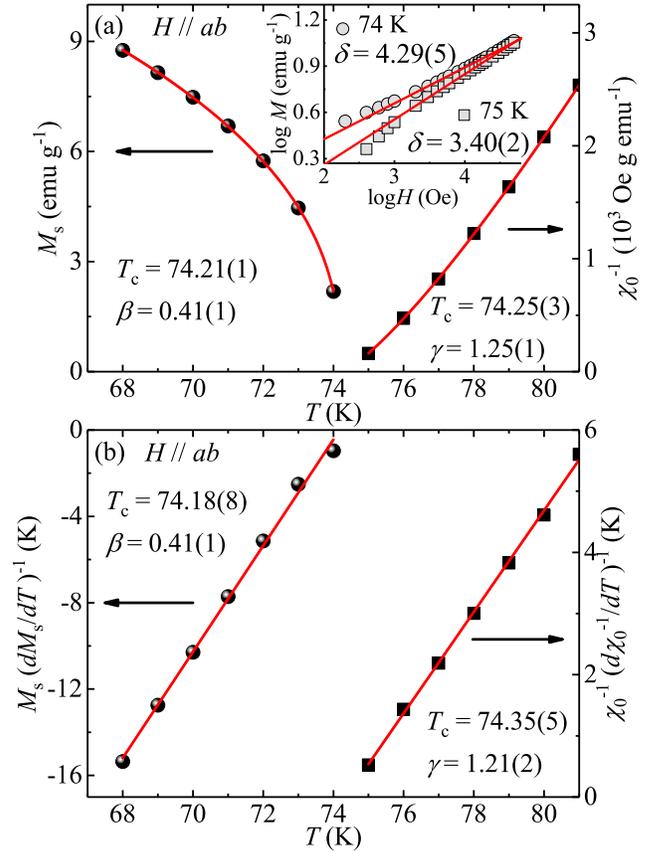}}
\caption{(Color online) (a) Temperature dependence of the spontaneous magnetization $M_s$ (left) and the inverse initial susceptibility $\chi_0^{-1}$ (right) with solid fitting curves. Inset shows log$M$ vs log$H$ collected at 74 and 75 K with linear fitting curves. (b) Kouvel-Fisher plots of $M_s(dM_s/dT)^{-1}$ (left) and $\chi_0^{-1}(d\chi_0^{-1}/dT)^{-1}$ (right) with solid fitting curves.}
\label{KF}
\end{figure}

From the Landau theory of phase transition, the Gibbs free energy $G$ for FM-paramagnetic(PM) transition can be expressed as
\begin{equation}
G(T,M)=G_0+aM^2+bM^4-MH,
\end{equation}
where the equilibrium magnetization $M$ is the order parameter, and the coefficients $a$ and $b$ are the temperature-dependent parameters. At equilibrium $\partial G/\partial M = 0$ (i.e., energy minimization) and the magnetic equation of state can be expressed as
\begin{equation}
H/M=2a+4bM^2.
\end{equation}
Thus, the Arrott plot of $M^2$ vs $H/M$ should appear as parallel straight lines for different temperatures above and below $T_c$ in the high field region.\cite{Arrott1} The intercepts of $M^2$ on the $H/M$ axis is negative or positive depending on phenomena below or above $T_c$ and the line at $T_c$ passes through the origin. In order to properly determine the $T_c$ as well as the critical exponents $\beta$, $\gamma$, and $\delta$, the modified Arrott plot with a self-consistent method was used.\cite{Kellner, Pramanik} Figure 3 presents the initial isotherms ranging from 5 to 90 K and the modified Arrott plot of $M^{1/\beta}$ vs $(H/M)^{1/\gamma}$ around $T_c$ for Mn$_3$Si$_2$Te$_6$. This gives $\chi_0^{-1}(T)$ and $M_s(T)$ as the intercepts on the $H/M$ axis and positive $M^2$ axis, respectively.

Figure 4(a) exhibits the final $M_s(T)$ and $\chi_0^{-1}(T)$ as a function of temperature. According to Eqs. (1) and (2), the critical exponents $\beta = 0.41(1)$ with $T_c = 74.21(1)$ K, and $\gamma = 1.25(1)$ with $T_c = 74.25(3)$ K, are obtained. In addition, there is also the Kouvel-Fisher (KF) relation,\cite{Kouvel}
\begin{equation}
M_s(T)[dM_s(T)/dT]^{-1} = (T-T_c)/\beta,
\end{equation}
\begin{equation}
\chi_0^{-1}(T)[d\chi_0^{-1}(T)/dT]^{-1} = (T-T_c)/\gamma.
\end{equation}
Linear fittings to the plots of $M_s(T)[dM_s(T)/dT]^{-1}$ and $\chi_0^{-1}(T)[d\chi_0^{-1}(T)/dT]^{-1}$ vs $T$ in Fig. 4(b) yield $\beta = 0.41(1)$ with $T_c = 74.18(8)$ K, and $\gamma = 1.21(2)$ with $T_c = 74.35(5)$ K. The third exponent $\delta$ can be calculated from the Widom scaling relation $\delta = 1+\gamma/\beta$. From $\beta$ and $\gamma$ obtained with the modified Arrott plot and the Kouvel-Fisher plot, $\delta$ = 4.05(5) and 3.95(2) are obtained, respectively, which are close to the direct fits of $\delta$ taking into account that $M = DH^{1/\delta}$ near $T_c$ [$\delta$ = 4.29(5) at 74 K and 3.40(2) at 75 K, inset in Fig. 4(a)].

\begin{figure}
\centerline{\includegraphics[scale=1]{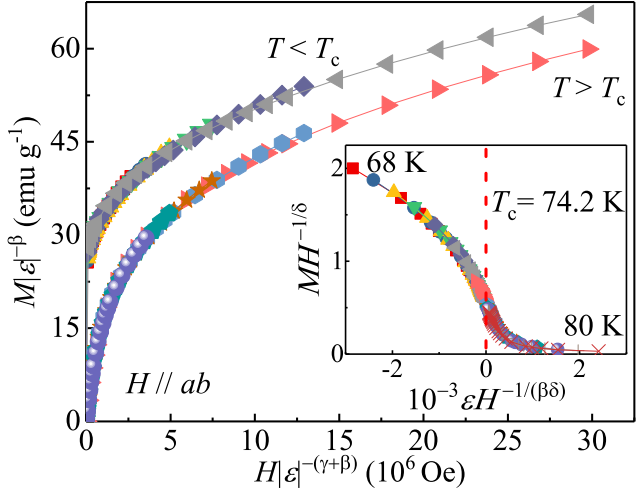}}
\caption{(Color online) Scaled magnetization $m$ vs scaled field $h$ below and above $T_c$ for Mn$_3$Si$_2$Te$_6$. Inset: the rescaling of the $M(H)$ curves by $MH^{-1/\delta}$ vs $\varepsilon H^{-1/(\beta\delta)}$.}
\label{renomalized}
\end{figure}

\begin{figure}
\centerline{\includegraphics[scale=1]{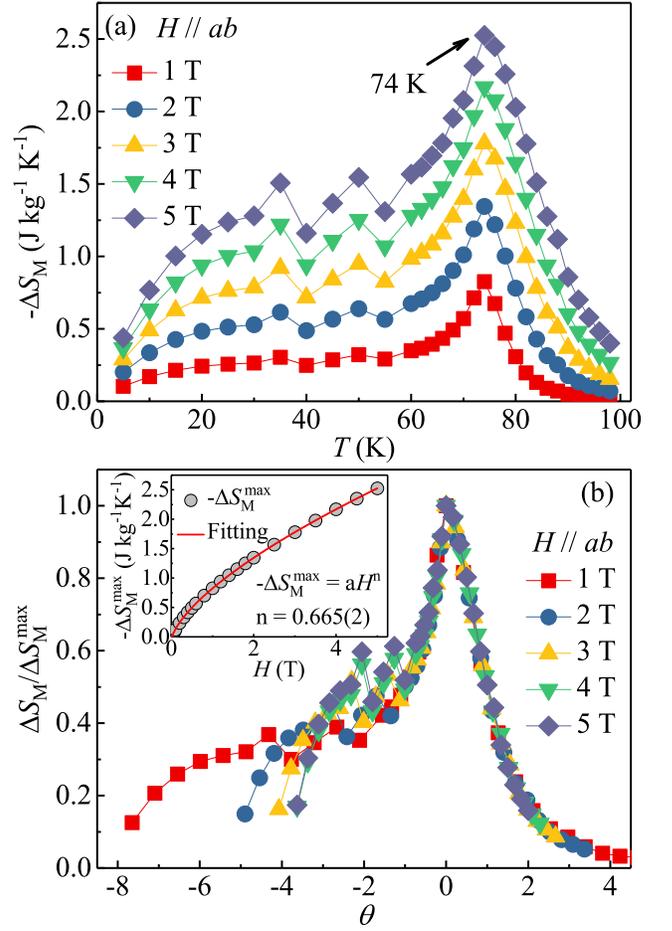}}
\caption{(Color online) (a) The magnetic entropy change $-\Delta S_M$ obtained from magnetization at various magnetic fields change in the $ab$ plane. (b) Normalized $\Delta S_M$ as a function of the rescaled temperature $\theta$. Inset: magnetic field dependence of the maximum magnetic entropy change $-\Delta S_M^{max}$ with power law fitting in red solid line.}
\label{renomalized}
\end{figure}

\begin{figure}
\centerline{\includegraphics[scale=1]{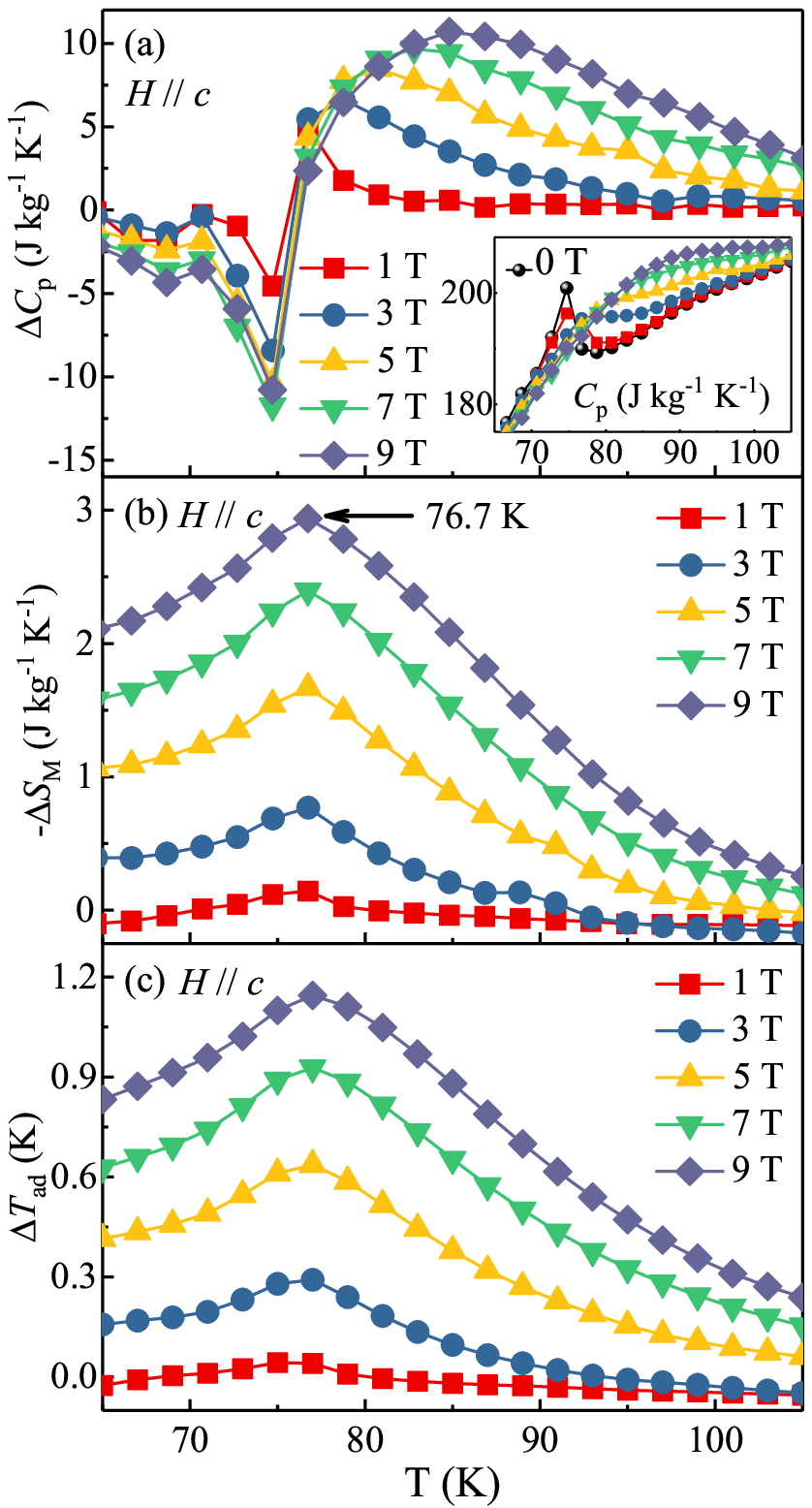}}
\caption{(Color online) Temperature dependences of (a) the specific heat change $\Delta C_p$, (b) the magnetic entropy change $-\Delta S_M$ and (c) the adiabatic temperature change $\Delta T_{ad}$ for Mn$_3$Si$_2$Te$_6$ at the indicated fields. Inset shows the temperature dependence of specific heat $C_p$.}
\label{renomalized}
\end{figure}

Scaling analysis can be used to estimate the reliability of the obtained critical exponents and $T_c$. From Eq. (5), scaled $m$ vs scaled $h$, all the data collapse on two separate branches below and above $T_c$, as depicted in Fig. 5. The scaling equation of state takes another form,
\begin{equation}
\frac{H}{M^\delta} = k\left(\frac{\varepsilon}{H^{1/\beta}}\right),
\end{equation}
where $k(x)$ is the scaling function. From Eq. (10), all the data should also fall into a single curve. This is indeed seen [inset in Fig. 5]; the $MH^{-1/\delta}$ vs $\varepsilon H^{-1/(\beta\delta)}$ experimental data for Mn$_3$Si$_2$Te$_6$ collapse into a single curve and the $T_c$ locates at the zero point of the horizontal axis. The well-rescaled curves further confirm the reliability of the obtained critical exponents.

Next, it is important to understand the nature as well as the range of interaction in this material. In a homogeneous magnet the universality class of the magnetic phase transition depends on the exchange distance $J(r)$. In renormalization group theory analysis the interaction decays with distance $r$ as
\begin{equation}
J(r) \approx r^{-(3+\sigma)},
\end{equation}
where $\sigma$ is a positive constant.\cite{Fisher1972} Moreover, the susceptibility exponent $\gamma$ is predicted as
\begin{multline}
\gamma = 1+\frac{4}{d}\left(\frac{n+2}{n+8}\right)\Delta\sigma+\frac{8(n+2)(n-4)}{d^2(n+8)^2}\\\times\left[1+\frac{2G(\frac{d}{2})(7n+20)}{(n-4)(n+8)}\right]\Delta\sigma^2,
\end{multline}
where $\Delta\sigma = (\sigma-\frac{d}{2})$ and $G(\frac{d}{2})=3-\frac{1}{4}(\frac{d}{2})^2$, $n$ is the spin dimensionality.\cite{Fischer} When $\sigma > 2$, the Heisenberg model is valid for 3D isotropic magnet, where $J(r)$ decreases faster than $r^{-5}$. When $\sigma \leq 3/2$, the mean-field model is satisfied, expecting that $J(r)$ decreases slower than $r^{-4.5}$. In the present case, $\sigma$ = 1.79, then the correlation length critical exponent $\nu$ = 0.676 ($\nu = \gamma/\sigma$), and $\alpha$ = -0.028 ($\alpha= 2 - \nu d$). It is found that the magnetic exchange distance decays as  $J(r)\approx r^{-4.79}$, which lies between that of 3D Heisenberg model and mean-field model.

Then we estimate its magnetic entropy change
\begin{equation}
\Delta S_M(T,H) = \int_0^H \left[\frac{\partial S(T,H)}{\partial H}\right]_TdH.
\end{equation}
With the Maxwell's relation $\left[\frac{\partial S(T,H)}{\partial H}\right]_T$ = $\left[\frac{\partial M(T,H)}{\partial T}\right]_H$, it can be further written as:\cite{Amaral}
\begin{equation}
\Delta S_M(T,H) = \int_0^H \left[\frac{\partial M(T,H)}{\partial T}\right]_HdH.
\end{equation}
In the case of magnetization measured at small discrete magnetic field and temperature intervals [Fig. 3(a)], $\Delta S_M(T,H)$ could be practically approximated as,
\begin{equation}
\Delta S_M(T_i,H) = \frac{\int_0^HM(T_i,H)dH-\int_0^HM(T_{i+1},H)dH}{T_i-T_{i+1}}.
\end{equation}
Figure 6(a) gives the calculated $-\Delta S_M$ as a function of temperature. All the $-\Delta S_M(T,H)$ curves present a pronounced peak at $T_c$, and the peak broads asymmetrically on both sides with increasing field. The maximum value of $-\Delta S_M$ reaches 2.53 J kg$^{-1}$ K$^{-1}$ with in-plane field change of 5 T. This is comparable to Mn$_{2-x}$Cr$_{x}$Sb but smaller than in MnFeP$_{0.45}$As$_{0.55}$ or Gd$_{5}$Ge$_{2}$Si$_{2}$ magnetic refrigerant materials.\cite{CaronL,TegusO}

Scaling analysis of $-\Delta S_M$ can be built by normalizing all the $-\Delta S_M$ curves against the respective maximum $-\Delta S_M^{max}$, namely, $\Delta S_M/\Delta S_M^{max}$ by rescaling the temperature $\theta$ as defined in the following equations,\cite{Franco}
\begin{equation}
\theta_- = (T_{peak}-T)/(T_{r1}-T_{peak}), T<T_{peak},
\end{equation}
\begin{equation}
\theta_+ = (T-T_{peak})/(T_{r2}-T_{peak}), T>T_{peak},
\end{equation}
where $T_{r1}$ and $T_{r2}$ are the temperatures of the two reference points that have been selected as those corresponding to $\Delta S_M(T_{r1},T_{r2}) = \frac{1}{2}\Delta S_M^{max}$. Following this method, all the $-\Delta S_M(T,H)$ curves in various fields collapse into a single curve in the vicinity of $T_c$ [Fig. 6(b)]. In the framework of the mean-field theory, $-\Delta S_M^{max} = -1.07qR(g\mu_BJH/k_BT_c)^{2/3} \propto H^{2/3}$, where $q$ is the number of magnetic ions, $R$ is the gas constant, and $g$ is the Lande factor.\cite{Oes} In fact, more universally, it should follow a power law relation, $-\Delta S_M^{max} = aH^n$, where $n$ depends on the magnetic state of the sample. Fitting of the field dependence of $-\Delta S_M^{max}$ with $\mathbf{H} // \mathbf{ab}$ gives $n = 0.665(2)$ [inset in Fig. 6(b)], close to the typical value of 2/3 within mean-field model.

Finally, we also estimate the $-\Delta S_M$ from heat capacity measurement with out-of-plane fields up to 9 T. The $\lambda$ peak observed at $T_c$ = 74.7 K in zero field [inset in Fig. 7(a)], corresponding well to the PM-FM transition, is gradually suppressed in fields. Figure 7(a) shows the calculated heat capacity change $\Delta C_p = C_p(T,H)-C_p(T,0)$ as a function of temperature in various fields. Obviously, $\Delta C_p < 0$ for $T < T_c$ and $\Delta C_p > 0$ for $T > T_c$, whilst, it changes sharply from negative to positive at $T_c$, corresponding to the change from FM to PM. The entropy $S(T,H)$ can be deduced by
\begin{equation}
S(T,H) = \int_0^T \frac{C_p(T,H)}{T}dT.
\end{equation}
Assuming the electronic and lattice contributions are not field dependent and in an adiabatic process of changing the field, the magnetic entropy change $-\Delta S_M$ can be straightly obtained $-\Delta S_M(T,H) = S_M(T,H)-S_M(T,0)$. The adiabatic temperature change $\Delta T_{ad}$ caused by the field change can be obtained by $\Delta T_{ad}(T,H) = T(S,H)-T(S,0)$, where $T(S,H)$ and $T(S,0)$ are the temperatures in the field $H \neq 0$ and $H = 0$, respectively, at constant total entropy $S(T,H)$. Figures 7(b) and 7(c) exhibit the temperature dependence of $-\Delta S_M$ and $\Delta T_{ad}$ estimated from heat capacity with out-of-plane field. The maxima of $-\Delta S_M$ and $\Delta T_{ad}$ increase with increase field and reach the values of 2.94 J kg$^{-1}$ K$^{-1}$ and 1.14 K, respectively, with the field change of 9 T.

\section{CONCLUSIONS}

In summary, we have studied the critical behavior and magnetocaloric effect around the FM-PM transition in Mn$_3$Si$_2$Te$_6$ single crystal. The ferrimagnetic transition in Mn$_3$Si$_2$Te$_6$ is identified to be second order in nature. The critical exponents $\beta$, $\gamma$, and $\delta$ estimated from various techniques match reasonably well and follow the scaling equation, suggesting a long-range magnetic interaction with the exchange distance decaying as $J(r)\approx r^{-4.79}$. Magnetocaloric effect is about one order of magnitude smaller when compared to other magnetorefrigerant candidate materials.

\section*{Acknowledgements}
This work was supported by the US DOE-BES, Division of Materials Science and Engineering, under Contract No. DE-SC0012704 (BNL).

\end{document}